\documentclass[letter,11pt]{article}
\pdfoutput=1

\usepackage{jheppub-mod} 

\usepackage[T1]{fontenc} 
\usepackage{amsthm}

\usepackage{listings}
\usepackage{mathrsfs}

\usepackage{xcolor} 

\usepackage[utf8]{inputenc}

\newcommand{\ket}[1]{| #1 \rangle}

\newcommand{\be}{\begin{equation}}
\newcommand{\ee}{\end{equation}}

\usepackage{color}
\definecolor{purple}{rgb}{0.5,0,0.5}

\newcommand{\es}[2] {\begin{equation} \label{#1} \begin{split} #2 \end{split} \end{equation}}
\def\le{\left}
\def\ri{\right}
\newcommand\ov{\over}

\newcommand\Lam{\Lambda}

\newcommand\Om{\Omega}
\newcommand\p{\ensuremath{\partial}}
\newcommand\abs[1]{\ensuremath{\left\lvert{#1}\right\rvert}}

\title{On the Entropy Cone for Large Regions at Late Times}

\author[a]{Ning Bao}
\author[b]{and M\'ark Mezei}
\affiliation[a]{Berkeley Center for Theoretical Physics,\\University of California, Berkeley, CA 94720, USA}
\affiliation[b]{Simons Center for Geometry and Physics,\\SUNY, Stony Brook, NY 11794, USA}

\emailAdd{ningbao75@gmail.com, mmezei@scgp.stonybrook.edu}

\abstract{In this note we show that the holographic entanglement entropy inequalities that hold for constant time slices are also valid for large regions and times in covariant spacetimes containing collapsing black branes, where the leading part of the entropy is computed by the simplified membrane theory of \cite{Mezei:2018jco}.  }

\begin{document} 
\maketitle
\flushbottom

\section{Introduction}

The study of holographic entanglement entropy via the Ryu-Takayanagi (RT) formula  \cite{Ryu:2006bv} has profoundly affected our understanding and interpretation of the AdS/CFT correspondence \cite{Maldacena:1997re}. In particular, it has been found that on constant time slices,  in addition to constraints obeyed by the entanglement entropies of all quantum systems, further constraints are true for holographic systems \cite{Hayden:2011ag, Bao:2015bfa}. These constraint inequalities serve as necessary conditions for what it means for a state to be holographically dual to a classical spacetime geometry.

 The result of  \cite{Ryu:2006bv} has also been extended to time-dependent situations by Hubeny, Rangamani, and Takayanagi (HRT) \cite{Hubeny:2007xt}. This work has also been usefully reformulated by Wall \cite{Wall:2012uf}, know as maximin. In the context of this latter maximin reformulation, the first of the new holographic entanglement entropy inequalities, the monogamy of mutual information, as proven in \cite{Hayden:2011ag}, was also proven for these time-dependent situations \cite{Wall:2012uf}.

These methods, however, were unable to prove the further inequalities found in \cite{Bao:2015bfa}, as they sensitively required the regions appearing on either side of the inequality to be non-overlapping. Because all of the further inequalities had overlapping regions, the question of whether or not these inequalities were true in time-dependent contexts remained an open one.

In recent work \cite{Mezei:2018jco}, however, it has been found that in certain limits the calculation of the HRT entanglement entropy can be reformulated into finding a codimension-1 minimal membrane with an angle dependent tension in flat space, where the bulk geometry is transformed into the membrane tension. In this work, we will argue that this is sufficient for proving that the further inequalities found in \cite{Bao:2015bfa} also hold in this limit of HRT.

The organization of the paper will be as follows. In section \ref{sec:ineqs}, we will provide a review of the proof methods in the RT case for the holographic entanglement entropy inequalities. In section \ref{sec:membrane}, we will describe the new methods of \cite{Mezei:2018jco}. In section \ref{sec:Inclusion}, we will argue for why the techniques for proving holographic entanglement entropies in the static case generalize to the HRT entropies in the limit in which the methods of \cite{Mezei:2018jco} apply.

\section{Holographic Entanglement Entropy Inequalities} \label{sec:ineqs}
\subsection{Inequalities from Ryu-Takayanagi}\label{sec:ineqs1}

\begin{figure}[!h]
\caption{Example of a holographic proof of strong subadditivity. The red and blue curves and cut and reglued at the point of intersection in the bulk to two green and maroon curves that nonminimally subtend $B$ and $ABC$. Note that here $A$, $B$, and $C$ have been chosen to be adjacent, single regions in a 2+1 dimensional bulk theory for simplicity only; inclusion-exclusion is able to generalize away from this in terms of adjacency, number of subregions, and spacetime dimension. \label{fig:SSA}}
\includegraphics[width=8cm]{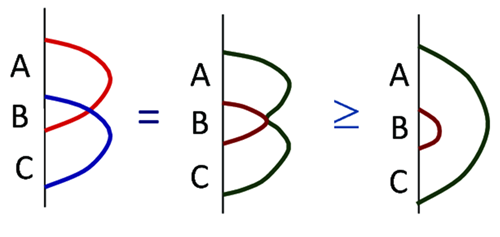}
\centering
\end{figure}

The Ryu-Takayanagi formula states that the entanglement entropy $S$ of a boundary subregion $X$ is given by the area $A_{\textnormal{min}}(X)$ of the minimal surface through the bulk  that is  homologous to the boundary region:
\begin{equation}
S(X)=\frac{A_{\textnormal{min}}(X)}{4G}\,.
\end{equation}
This equation geometrizes the problem of calculating the entanglement entropy in a way that makes computations and proofs with respect to entanglement entropies more tractable. Consider, for example, three boundary subregions $A$, $B$, and $C$. These regions are allowed to be adjacent or not, and either single regions or disjoint unions of multiple regions. It is straightforward to show that the minimal surface subtending $AB$ and the minimal surface subtending $BC$ can be cut and reglued to non-minimally subtend $B$ and $ABC$ (see Figure \ref{fig:SSA}). This generic proof method, of cutting and regluing unions of minimal surfaces corresponding to the left-hand side of an inequality to non-minimally cover the full set of regions on the right hand side of the inequality, is called inclusion-exclusion. Because the areas of the minimal surfaces give the entanglement entropies, this proves strong subaddivity,
\begin{equation}
S(AB)+S(BC)\geq S(A)+S(ABC)
\end{equation}
on a constant time slice in a holographic spacetime. 

 Note also that this geometrization places an additional constraint that is not true of all entanglement entropies,\footnote{Namely, this is the equivalence of entanglement entropy and bulk homologous minimal surface area.} and thus can lead to more constraining inequalities that exclude certain quantum states. The first inequality of this type discovered by \cite{Hayden:2011ag} was the monogamy of mutual information (MMI):
 \begin{equation}
 S(AB)+S(BC)+S(AC)\geq S(A)+S(B)+S(C)+S(ABC)\,.
 \end{equation}
 
 This inequality can clearly also be proven using inclusion-exclusion, as above. A simple example of a quantum state which does not satisfy this inequality is the 4-body GHZ state, ${1\ov \sqrt{2}}\le(\ket{0000}+\ket{1111}\ri)$,\footnote{While in this example the bodies in question are single qubit system, one can easily generalize this entanglement pattern to much larger Hilbert space dimensions.} thus demonstrating that this inequality, unlike strong subadditivity, excludes certain quantum states from being dual to classical bulk spacetimes.
 
 Following the inclusion-exclusion proof method, more holographic entanglement entropy inequalities were proven for constant time slices in holographic spacetimes in \cite{Bao:2015bfa}, including an infinite family generalizing the MMI inequality. We refer the reader to \cite{Bao:2015bfa} for more details, but for 5-body mixed state systems (where each ``body'' is a non-overlapping boundary subregions, with a sixth region consisting of the rest of the boundary that purifies the density matrix of the union of the five non-overlapping boundary subregions) the known new inequalities for constant time slices are:
 \begin{itemize}
\item $S(ABC)+S(BCD)+S(CDE)+S(DEA)+S(EAB) \geq S(AB)+S(BC)+S(CD)+S(DE)+S(EA)+S(ABCDE)\,,$
\item $2S(ABC) + S(ABD) + S(ABE) + S(ACD) + S(ADE) + S(BCE) + S(BDE) \geq S(AB) + S(ABCD) + S(ABCE) + S(ABDE) + S(AC) + S(AD) + S(BC) + S(BE) + S(DE)\,,$
\item $S(ABE) + S(ABC) + S(ABD) + S(ACD) + S(ACE) + S(ADE) + S(BCE) + S(BDE) + S(CDE) \geq S(AB) + S(ABCE) + S(ABDE) + S(AC) + S(ACDE) + S(AD) + S(BCD) + S(BE) + S(CE) + S(DE)\,,$
\item $S(ABC) + S(ABD) + S(ABE) + S(ACD) + S(ACE) + S(BC) + S(DE) \geq S(AB) + S(ABCD) + S(ABCE) + S(AC) + S(ADE) + S(B) + S(C) + S(D) + S(E)\,,$
\item $3S(ABC) + 3S(ABD) + 3S(ACE) + S(ABE) + S(ACD) + S(ADE) + S(BCD) + S(BCE) + S(BDE) + S(CDE) \geq 2S(AB) + 2S(ABCD) + 2S(ABCE) + 2S(AC) + 2S(BD) + 2S(CE) + S(ABDE) + S(ACDE) + S(AD) + S(AE) + S(BC) + S(DE)\,.$
\end{itemize}

We remark that these proofs break down for higher derivative gravity theories, where the entanglement entropy is computed by minimizing a generalized area functional containing higher derivative terms, such as the extrinsic curvature \cite{Hung:2011xb,Dong:2013qoa,Camps:2013zua}. The problem arises in the ``recoloring" step of Figure \ref{fig:SSA}: the two generalized areas do not necessarily agree and there can be large contributions from cusps.

\subsection{Inequalities from HRT}

It is a natural question to ask how the RT formula generalizes to covariant cases. In this case, the HRT formula \cite{Hubeny:2007xt} gives that the entanglement entropy $S(X)$ of a boundary region $X$ is given by the area of the extremal surface $A_{\textnormal{ext}}(X)$ homologous to $X$:
\begin{equation}
S(X)=\frac{A_{\textnormal{ext}}(X)}{4G}\,.
\end{equation}
The HRT proposal was reformulated by Wall in \cite{Wall:2012uf}, in which a ``maximin'' prescription for calculating the area of the extremal surface was proposed and proven equivalent; for a detailed review of this proposal, we refer the interested reader to \cite{Wall:2012uf}, as we will focus here on only a few key ingredients thereof.

One can use maximin to prove holographic entanglement entropy inequalities in covariant situations. Famously, in \cite{Wall:2012uf} MMI was proven to also hold covariantly. There are two main steps required in a maximin proof of an entanglement entropy inequality: first, one must find a single Cauchy slice on which minimal surfaces homologous to all of the regions listed in the smaller side of the inequality can be simultaneously lain. Once this has been done, one can simply proceed with inclusion-exclusion as before. It can be trivially seen that this can therefore be done for MMI.

This proof method notably fails when the regions on the smaller side of the inequality have nontrivial overlap, e.g. are not all mutually disjoint or nested. As all of the further holographic entanglement entropy inequalities beyond MMI have this property, it seems difficult to proceed further with the maximin approach, something which has been studied in detail in \cite{Rota:2017ubr}.

\section{A Limit of HRT}\label{sec:membrane}

Let us consider the following nonequilibrium setup: we take a sparsely entangled initial state with finite energy density and a narrow energy distribution and let this state evolve unitarily. A natural way to prepare such a state is a quantum quench, where we start with the vacuum state of a local Hamiltonian and then abruptly change the Hamiltionan and let the system unitarily evolve subsequently. In a chaotic system such a state undergoes local thermalization in time $t_\text{loc}$, and thereafter is governed by hydrodynamics. In a theory with a gravity dual a sparsely entangled initial state is dual to some matter distribution near the AdS boundary, which subsequently collapses to form a black brane -- this process is the dual of local thermalization. This collapse is fast, $t_\text{loc}\sim\beta$, where $\beta$ is the inverse temperature  associated to the state. The subsequent time evolution is described by fluid/gravity spacetimes dual to theories of hydrodynamics \cite{Janik:2005zt,Bhattacharyya:2008jc}.

We want to understand entanglement entropy in such states for large regions with characteristic size $R$ and at late times:
\es{Scaling}{
R,t\gg t_\text{loc}\,.
}
We focus on the homogeneous case below, where hydrodynamics is trivial. In the limit \eqref{Scaling} the entropy is expected to obey the scaling form:
\es{ScalingEE}{
S(A(t))=s_\text{th}(\beta) R^{d-1}\, f_A\le(t\ov R\ri)+\dots\,,
}
where $s_\text{th}(\beta)$ is the thermal entropy density and $f_A\le(t\ov R\ri)$ is a functional of $A$. We will refer to the leading piece of the entropy as the extensive piece. Note that an area law piece appears only at subleading order in the  $t_\text{loc}/R$ (or $\beta/R$) expansion. 

In \cite{Mezei:2018jco} it was shown that the extensive piece of the entropy is determined by an auxiliary membrane minimization problem. We continue to focus on the homogeneous case with constant $t$ Cauchy slices, and comment on generalizations later. Let us take a slab of Minkowski spacetime of height $t$, where the lower face of the slab represents the initial sparsely entangled state and the upper face is the time slice on which we want to compute the entropy. Let us consider a minimal energy membrane stretching between the two faces anchored on the entangling region $\p A(t)$ on the upper face on one end, ending perpendicularly on the lower face on the other, and governed by the action that equals the entropy on shell:
\es{AreaFunctScaled2}{
S(A(t))&=s_\text{th}  \int_0^t d\text{Area}\ { {\cal E}\le(v\ri)\ov \sqrt{1-v^2}}\,,\\
v&\equiv {n_t\ov\sqrt{ 1+ n_t^2}}\,,
 }
where $n_\mu$ is the local unit normal of the surface, and $n_t$ is its time component, $v$ is a local angle of the membrane with the vertical direction, and ${\cal E}\le(v\ri)$ (the membrane tension function) is an even function of $v$ that is determined by the equilibrium black brane geometry. For $d=2$ the membrane is a worldline with $v$ its speed. In higher dimensions it is a speed of a wave front obtained by considering the cross section of the membrane and is limited to be $\abs{v}\leq 1$. However, we also allow for the membrane to have horizontal segments (formally corresponding to $v=\infty$), corresponding to the replacement $ {{\cal E}\le(v\ri)\ov \sqrt{1-v^2}}\big\vert_{v=\infty}\to 1$. 

To follow our argument about the proof of entropy inequalities, one only needs to know that to determine the extensive part of the entropy, instead of solving an extremization problem for a codimension-2 surface, we can solve a minimization problem for a codimension-1 surface (the membrane) with the local action given in \eqref{AreaFunctScaled2}. We note that the membrane theory determination of the entropy is straightforwardly generalized to regions which do not lie at constant time: we simply have to modify the slab of Minkowski spacetime to have the appropriate Cauchy slice as its (wiggly) boundary.\footnote{It is clear that the answer will only depend on the entangling surface $\p A(t)$ and not the Cauchy slice that the entangling surface is on.} According to preliminary results \cite{inpreparation}, the membrane theory generalizes for fluid/gravity spacetimes in a way that the locality of the action is preserved, and the membrane is sensitive to the local fluid velocity and temperature. The membrane theory can also be generalized to include subleading corrections in $\beta/R$ \cite{inpreparation}; while the locality of the action is preserved, higher derivative terms, such as the extrinsic curvature (given by the derivative of the normal vector $n_\mu$) do appear. These features of the action will play a crucial role in the discussion of Section~\ref{sec:Inclusion}. 

The membrane phenomenology of entropy dynamics extends beyond holography, in  \cite{Nahum:2016muy,Jonay:2018yei} it was shown that the membrane theory can be derived for random quantum circuits, and in \cite{Jonay:2018yei} numerical evidence was presented for its applicability for a spin chain with a chaotic Hamiltonian. Based on these results and holography it was conjectured in  \cite{Jonay:2018yei,Mezei:2018jco} that the membrane theory describes entanglement entropy in the limit \eqref{Scaling} in all chaotic theories in states discussed above, hence it is proposed to have a comparably wide applicability to hydrodynamics.

For completeness, below we review the holographic derivation of the membrane theory \eqref{AreaFunctScaled2} for homogeneous spacetimes. Readers who are eager to see the concluding steps of our argument for the applicability of the entropy cone inequalities should jump directly to Section~\ref{sec:Inclusion}. 

\begin{figure}[!h]
\caption{Penrose diagram for two examples of time dependent spacetimes, for which our considerations apply. We have drawn an HRT surface that computes the entanglement entropy of a subregion at time $t$. {\bf Left:} Matter falling in from the boundary of AdS to form a black brane. The resulting spacetime can be thought of consisting of three regions, a pure AdS part, a nonequilibrium region (drawn in green), and the equilibrium black brane at late times. The Vaidya spacetimes studied in \cite{Liu:2013iza,Liu:2013qca} fall in this class of spacetimes. We divide the HRT surface is divided in three parts, and only the second contributes to the extensive part of the entropy and it is determined by the membrane theory \eqref{AreaFunctScaled2}. {\bf Right:} In this example a short-range entangled initial state is prepared from a boundary state $\ket{B}$ with a Euclidean path integral \cite{Hartman:2013qma}, and the dual spacetime is an eternal black hole cut in half by and end of the world brane (drawn by green dashed line). In this figure the third part of the HRT surface is missing.   \label{fig:HRT}}
\vspace{0.3cm}
\includegraphics[width=6cm]{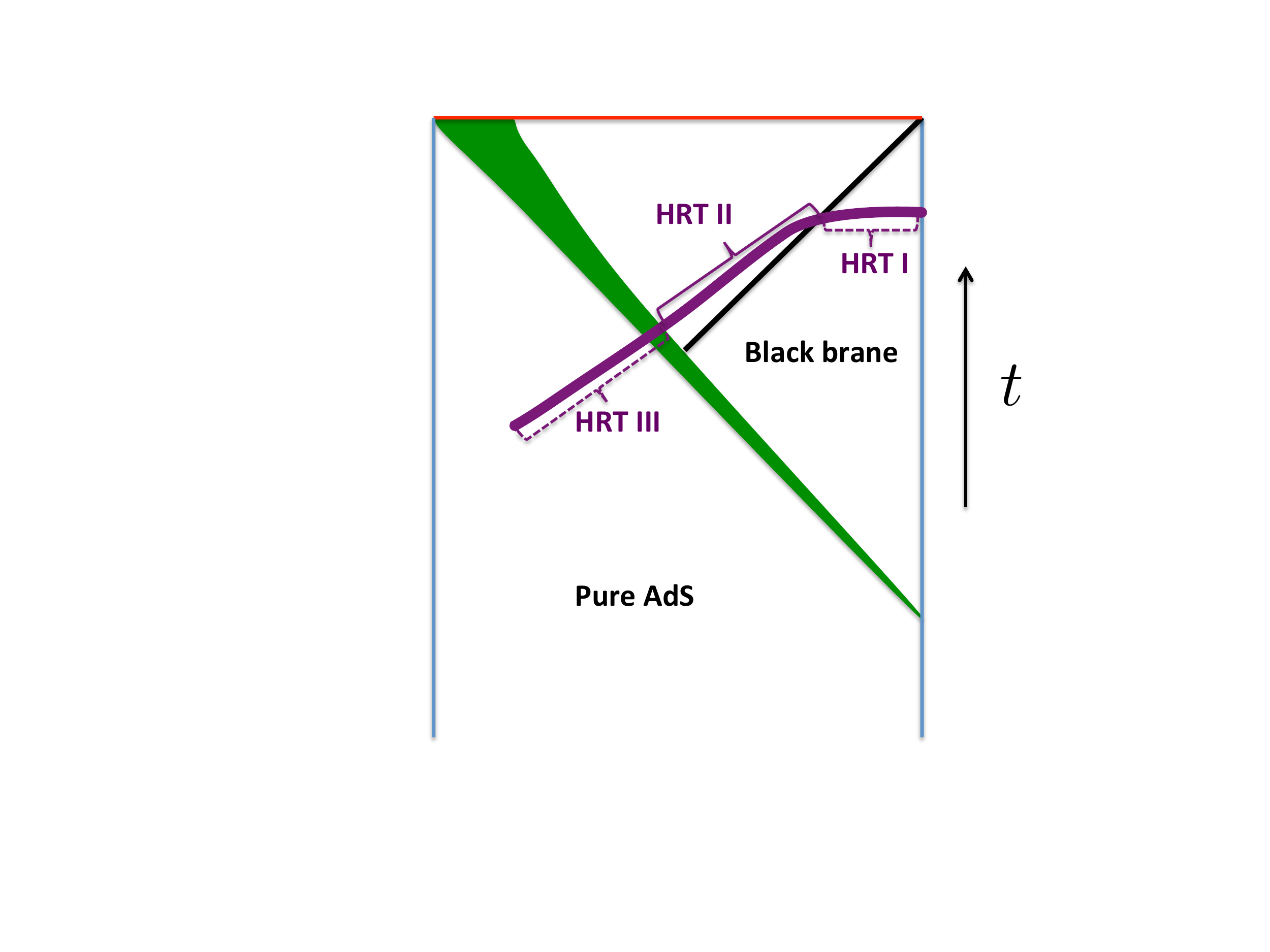}\hspace{1cm}
\includegraphics[width=7cm]{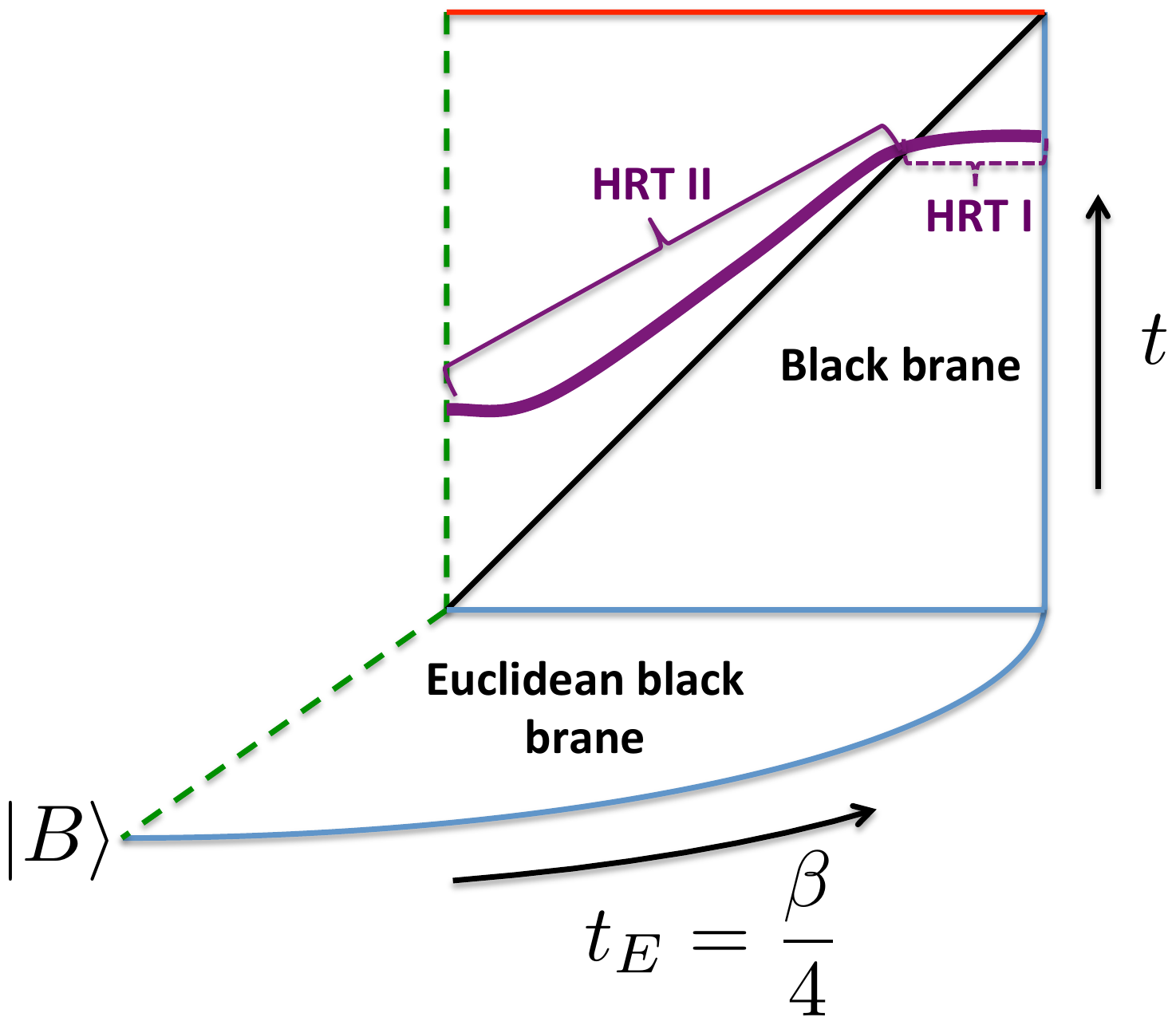}
\centering
\end{figure}

There are two steps in the derivation of the membrane theory. First, we decompose the HRT surface into three parts: the part that lives outside the black brane horizon and connects to the AdS boundary, the second that lives at and behind the horizon of the equilibrium black brane, and the last that lives in the nonequilibrium part of the spacetime and in the region of spacetime that represents the short range entangled initial state, see Figure \ref{fig:HRT}. Only the second part contributes to the extensive part of the entropy \cite{Hartman:2013qma,Liu:2013iza,Liu:2013qca,Mezei:2016zxg}, and we can compute its shape in a scaling limit \cite{Mezei:2016zxg,Mezei:2018jco} (to the required precision).\footnote{To understand the shape of the other parts, one in fact needs to apply different scalings, but we do not consider these other parts in what follows.} Let us comment on the other two parts briefly. While we have not worked out the details completely, we expect the first part to be of the shape of a tube (with cross section $\p A$, length along the $z$ direction, and not moving in Schwarzschild time to leading order in $\beta/R$), which gives an exact area  law for the entropy which is down by $\beta/R$ compared to the extensive contribution. We remark that an exact area law trivially saturates all holographic inequalities \cite{Bao:2015boa}. The second part of the HRT surface contributes at every order of the $\beta/R$ expansion, and we have commented above on the structure of these contributions. Finally, the third part depends sensitively on the protocol that prepared the initial sparsely entangled state. For the example in the left of Figure \ref{fig:HRT}, the third part consists of a part that crosses the infalling matter shell (and is again expected to be tube-shaped) and a part that is in the pure AdS region of the spacetime, both giving subleading in  $\beta/R$ contributions to the entropy, while in the example on the right of Figure \ref{fig:HRT}, the word brane quench model of \cite{Hartman:2013qma} the third part is absent \cite{Mezei:2016zxg}.\footnote{See \cite{Kourkoulou:2017zaj,Almheiri:2018ijj,Cooper:2018cmb} for recent interesting discussions of these geometries.}

The second step involves setting up coordinates and implementing the scaling limit. Because this part of the  HRT surface lives in a static black brane background (see Figure \ref{fig:HRT}), we can write the spacetime as
 \es{MetricBH}{
ds^2&= {1\ov z^2}\le[-a(z)dt^2-{2\ov b(z)}\,dtdz+ dr^2 +r^2 d\Om_{d-2}^2\ri]\,,
}
where the boundary is at radial position $z=0$, $(r,\Om)$ are boundary polar coordinates, and $t$ is the infalling time, which on the boundary is simply the Minkowski time. Without loss of generality we can set the horizon to be at $z_h=1$ (implying $a(1)=0$), and AdS  
asymptotics require that $a(0)=b(0)=1$. The HRT surface is codimension-2, hence we have to pick two dependent coordinates, which we choose to be $(r,z)$, followed by implementating a rescaling with $\Lam$, a large parameter, which we will eventually set to $1$:
\es{ScalingCoord}{
t&\equiv \Lam\,\hat{t}\,, \quad\hspace{-0.1cm} r(t,\Om)  \equiv \Lam \, \hat{r}(\hat{t},\Om) \,,\quad \hspace{-0.1cm}
 z(t,\Om)  \equiv \hat{z}(\hat{t},\Om)\,.
}
In the limit $\Lam\to \infty$ (corresponding to large $R/\beta$) the HRT area functional simplifies and takes the form in the scaled variables:
\es{AreaFunctScaled}{
S&=s_\text{th} \Lam^{d-1} \int d\hat{t} d\Om\ {\hat{r}^{d-2}\ov \hat{z}^{d-1}}\sqrt{Q}\,,\\
Q&\equiv (\p_{\hat{t}} \hat{r})^2- a(\hat{z})\le(1+{(\p_\Om \hat{r})^2\ov \hat{r}^2}\ri)\,.
} 
From now on we set $\Lam=1$ and drop the hats. We proceed by solving the $z$ equation of motion algebraically\footnote{For a proof of the invertibility of $c(z)$ see  \cite{Mezei:2018jco}. }
\es{ZEOM}{
z=c^{-1}(v^2)\,,
}
where 
\es{QuantDef}{
c(z)\equiv a(z)-{z a'(z)\ov 2(d-1)}\,, \qquad v\equiv {\p_t r\ov \sqrt{1+{(\p_\Om r)^2\ov r^2}}}\,.
}
Plugging the solution \eqref{ZEOM}  back into the action \eqref{AreaFunctScaled} we obtain the membrane theory \eqref{AreaFunctScaled2} in coordinates $(t,r,\Om)$ with $r(t,\Om)$ chosen as the dependent coordinate. It is a simple calculation to show that the geometric definition of $v$ given in \eqref{AreaFunctScaled2} equals the formula given in \eqref{QuantDef} in these coordinates.\footnote{Note that in these coordinates
\es{nExpr}{
n_\mu={1\ov \sqrt{1+{(\p_\Om r)^2\ov r^2}-(\p_t r)^2}}\le(\p_t r,\, -1,\,\p_\Om r\ri)\,;
}
simple algebra then proves the equivalence of the two formulas.} The membrane tension function is given by the holographic formula:
 \es{MembraneTension}{
 {\cal E}\le(v\ri)&={\sqrt{- a'(z)\ov 2(d-1)z^{2d-3}}}\Bigg\vert_{z=c^{-1}(v^2)}\,.
} 
An example membrane tension function for a Schwarzschild black brane is $ {\cal E}\le(v\ri)={v_E\ov (1-v^2)^{(d-2)/(2d)}}$.\footnote{The geometry is given by $a(z)=1-z^d,\, b(z)=1$ and $v_E={\le(d-2\ov d\ri)^{(d-2)/(2d)}\ov\le(2(d-1)\ov d\ri)^{(d-1)/d}}$.} Once the minimal membrane shape is determined (in polar coordinates $r(t,\Om)$ and from this $v(t,\Om)$) we can plug back into \eqref{ZEOM} to obtain the full HRT surface in the spacetime \eqref{MetricBH}. 

Let us discuss what the membrane theory has achieved.  We emphasize that HRT surfaces anchored on different shapes (even at the same time $t$) live on different Cauchy slices as we commented above, and this hampers the proof of many-party inequalities using the maximin approach. The membrane theory circumvents this problem by repackaging the entire spacetime \eqref{MetricBH} (and hence all possible Cauchy surfaces) into an angle dependent membrane tension function \eqref{MembraneTension}. All membranes live in the boundary Minkowski space, their energy determines the extensive part of the entropy, and once the minimal membrane is found we can reconstruct the HRT surface from the knowledge of the local value of $v$.

\section{Inclusion-Exclusion for a Limit of HRT}\label{sec:Inclusion}

A key ingredient of the construction in the previous section is that it converted the extremization problem of covariant holographic entanglement entropy to a geometric minimization problem. This should, once again, allow the utilization of inclusion-exclusion as a proof method. Moreover, this seems to allow for the sidestepping of the issue in \cite{Wall:2012uf}, as nowhere in this discussion is the mutual nesting or exclusion of regions on the smaller side of the inequalities needed.

In some sense, there is nothing intrinsically holographic about inclusion-exclusion. In fact, it was used to prove that the same set of inequalities for holographic entanglement entropy, indeed, are obeyed by condensed matter systems with area laws in \cite{Bao:2015boa}. The membrane theory is conjectured to describe the entanglement dynamics in the large region, late time limit  \eqref{Scaling} of all chaotic systems. In all these systems, our proof goes through, and assuming the validity of the membrane theory implies that  the static entropy cone inequalities are valid at late times for large subsystems of all chaotic systems in a quench setup. All that is needed is that the entanglement entropy is proportional to a partionable geometric minimization; at this point, the cutting and regluing can then simply be done in that geometry; it is not critical that that geometry is AdS space.

This means that the minimization procedure described in the previous section is sufficient to show that all of the holographic entropy inequalities thus far found for constant time slices also hold in this limit of HRT, as inclusion-exclusion holds in this case. Let us note, that similarly to the higher derivative gravity case discussed at the end of Section \ref{sec:ineqs1}, at subleading order in the $\beta/R$ expansion of the membrane theory, higher derivative terms appear in the action  \cite{inpreparation}, which prevent the proof of inequalities using the strategy summarized in Figure \ref{fig:SSA}. It would be interesting to understand, if there are interesting situations when an inequality is saturated for the extensive piece and one has to investigate subleading pieces. A trivial example when saturation happens is at very late times, when $S(A(t))=s_\text{th}\, {\rm vol}(A)+\dots$, and the extensive piece trivially saturates all inequalities provable using inclusion-exclusion.

\section*{Acknowledgments}
M\'ark Mezei would like to thank Julio Virrueta for collaboration on \cite{inpreparation}.
Ning Bao is supported by the National Science Foundation under grant number 82248-13067-44-PHPXH and by the Department of Energy under grant number DE-SC0019380. M\'ark Mezei is supported by the Simons Center for Geometry and Physics.

\bibliographystyle{utphys-modified}
\bibliography{Allrefs}

\end{document}